\documentclass{article}
\usepackage[preprint]{spconf}
\copyrightnotice{978-1-6654-7189-3/22/\$31.00~\copyright2023 IEEE}

\usepackage{amsmath,amsfonts}
\usepackage{algorithm}
\usepackage{array}
\usepackage[caption=false,font=normalsize,labelfont=sf,textfont=sf]{subfig}
\usepackage{textcomp}
\usepackage{stfloats}
\usepackage{url}
\usepackage{verbatim}
\usepackage{graphicx}
\usepackage{cite}

\usepackage{amsthm,amsopn}
\usepackage{bm}
\usepackage{hyperref}
\usepackage{microtype}
\usepackage{url}
\usepackage[capitalise]{cleveref}
\usepackage{graphicx,caption,lipsum}
\usepackage{booktabs,multirow}
\usepackage{placeins}
\usepackage{algpseudocode,algorithm}
\usepackage{color}
\usepackage[table,xcdraw]{xcolor}
\definecolor{Gray}{gray}{0.9}
\definecolor{brightturquoise}{rgb}{0.85, 1, 1}
\usepackage{bbm}
\usepackage{enumerate}   
\usepackage{adjustbox}
\usepackage[normalem]{ulem}
\usepackage{fixltx2e}
\usepackage{wrapfig}
\usepackage{threeparttable}
\usepackage{tikz}
\usepackage{booktabs}

\usepackage{seqsplit}

\title{Unconstrained Dysfluency Modeling for Dysfluent Speech Transcription and Detection}
\name{%
  \begin{tabular}{c}
    Jiachen Lian$^1$, Carly Feng$^{*1}$, \thanks{${\star}$ Work done when Carly was at UC Berkeley before joining Amazon}  Naasir Farooqi$^{1}$, Steve Li$^{2}$, Anshul Kashyap$^{1}$, \\
    Cheol Jun Cho$^{1}$, Peter Wu$^{1}$, Robbie Netzorg$^{1}$, Tingle Li$^{1}$, Gopala Krishna Anumanchipalli$^{1}$%
  \end{tabular}%
}

\address{$^{1}$ UC Berkeley $^{2}$ Harvard University \\
\tt \{jiachenlian, gopala\}@berkeley.edu}

\begin{document}
\ninept
\maketitle

\newcommand{\ma}[1]{{\color{orange}#1}}
\newcommand{\wh}[1]{{\color{magenta}#1}}
\newcommand{\jl}[1]{{\color{brown}#1}}

\newcommand{\nam}{AV-data2vec}

\newcommand{\overbar}[1]{\mkern 1.5mu\overline{\mkern-4.5mu#1\mkern-4.5mu}\mkern 1.5mu}

\begin{abstract}
Dysfluent speech modeling requires time-accurate and silence-aware transcription at both the word-level and phonetic-level. However, current research in dysfluency modeling primarily focuses on either transcription or detection, and the performance of each aspect remains limited. In this work, we present an unconstrained dysfluency modeling (UDM) approach that addresses both transcription and detection in an automatic and hierarchical manner. UDM eliminates the need for extensive manual annotation by providing a comprehensive solution. Furthermore, we introduce a simulated dysfluent dataset called VCTK$^{++}$ to enhance the capabilities of UDM in phonetic transcription. Our experimental results demonstrate the effectiveness and robustness of our proposed methods in both transcription and detection tasks.
\end{abstract}
\noindent\textbf{Index Terms}: dysfluent speech, transcription, detection
\section{Introduction}
In the field of speech analysis, a clear and universally agreed-upon definition of dysfluent speech is yet to be established. Dysfluencies are commonly associated with speech disorders such as stuttering, aphasia~\cite{brady2016aphasia}, and dyslexia~\cite{snowling2013dyslexia}, characterized by disruptions in the normal flow of speech~\cite{palfy2012pattern-dysfluent}. However, dysfluency also appears in normal conversational speech or spontaneous speech~\cite{pitt2005buckeye}, where individuals may experience hesitations or interruptions while speaking. In this context, we refer to dysfluent speech as any form of speech characterized by abnormal patterns such as repetition, prolongation, and irregular pauses, as discussed in~\cite{kouzelis2023weakly}. Within the domain of \textit{dysfluent speech modeling}, there is also a lack of a unified paradigm. Research efforts in this area can be roughly categorized as \textit{transcription} and \textit{detection}. 

\begin{figure*}[ht]
    \centering
    \includegraphics[height=10cm]{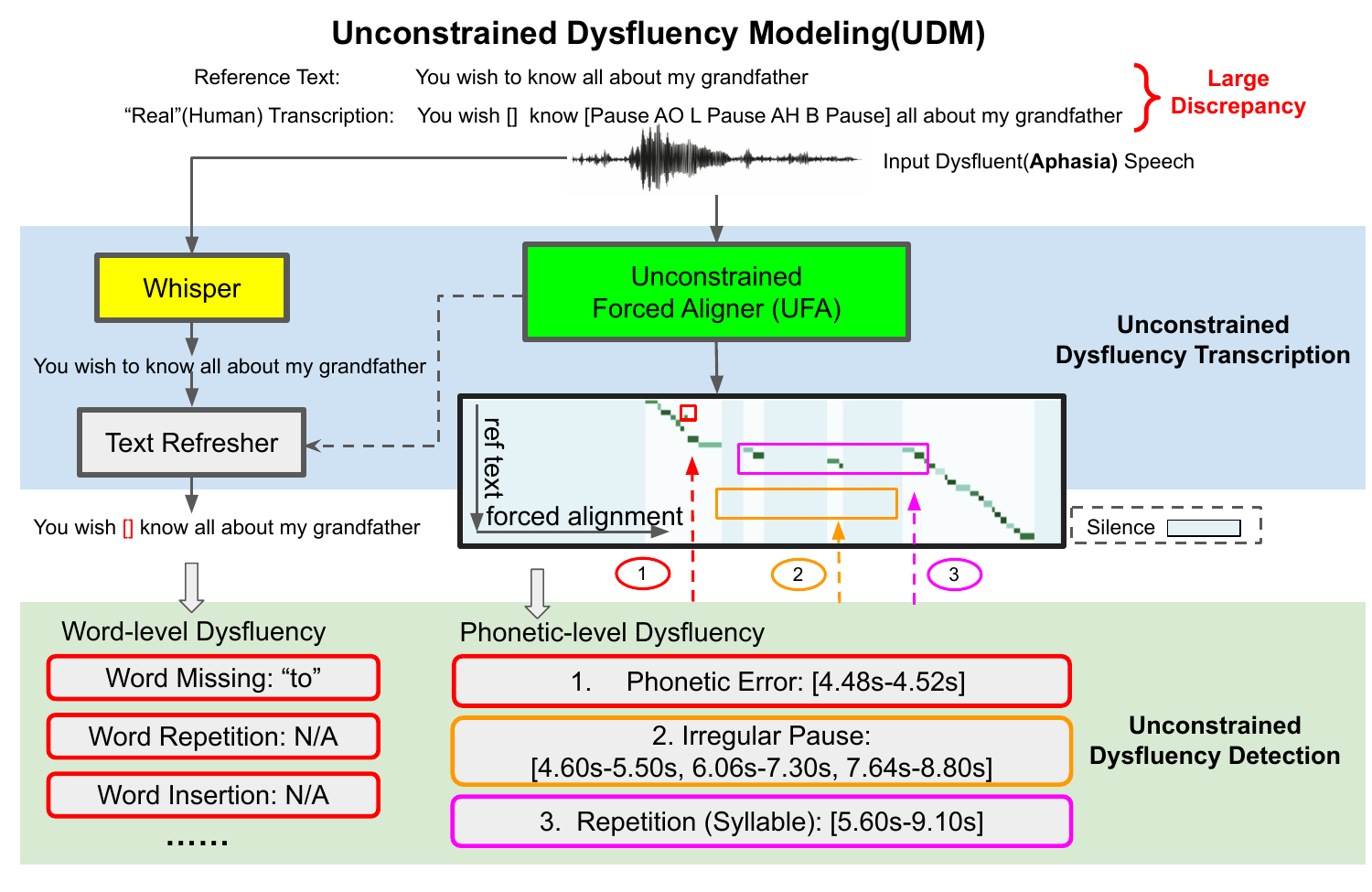}
    \caption[Unconstrained Dysfluency Modeling (Transcription and Detection) for aphasia speech example]{Unconstrained Dysfluency Modeling (Transcription and Detection) for aphasia speech instance. Here is an example of aphasia speech. The reference text is "You wish to know all about my grandfather," while the human transcription or ground truth differs significantly from the reference. Whisper recognizes it as perfect speech, while UFA is able to capture most of the dysfluency patterns. An audio sample of this can be found here\protect\footnotemark.}
    \label{UDM}
\end{figure*}

Current state-of-the-art word transcription models~\cite{radford2022whisper, lian2023av-data2vec, zhang2023google-usm, pratap2023scaling-speech, aghajanyan2023scaling-speech} and phonetic transcription models~\cite{wavlm-ctc} often struggle to accurately transcribe dysfluent speech. As a result, human annotations are still required and transcribers commonly resort to a two-step process. They first obtain automatic transcriptions and then manually annotate the missing portions, a laborious and time-consuming task that is comparable to manual annotation from scratch. 
WhisperX\cite{bain2023whisperx} recently extends Whisper~\cite{radford2022whisper} by generating timestamps for individual words. However, it still delivers limited performance for dysfluent speech. Moreover, time-accurate \textit{phonetic} transcriptions might be a better representation to capture various dysfluency types. Another requirement is that phonetic transcriptions should be sensitive to silences as it might indicate a block or poor breath-speech coordination. \cite{kouzelis2023weakly} recently proposed a time-accurate and silence-aware neural forced aligner, where a weighted finite-state transducer (WFST) is introduced for modeling dysfluency patterns such as repetition. However, this approach assumes that there is minimal deviation between the reference and "real" transcribed text. In real-life dysfluent speech, such as the example shown in Figure \ref{UDM}, this assumption may not hold true. 

Research on dysfluency detection has traditionally been conducted independently of dysfluency transcription and has recently been dominated by end-to-end methods. These approaches typically focus on either utterance-level detection \cite{kourkounakis2021fluentnet, alharbi2017segment-detection2, alharbi2020segment-detection3, segment-detection4}, or frame-level detection \cite{harvill2022frame-level-stutter, shonibare2022frame-detection2}. However, these studies primarily address data-driven classification problems and do not explicitly incorporate dysfluency transcription into their detection methods. A unified framework that integrates dysfluency transcription and detection is essential to develop an efficient and robust dysfluency modeling system.

In this study, we propose an \textit{unconstrained dysfluency modeling} (UDM) approach that integrates dysfluent speech transcription and pattern detection in an automatic manner with no human effort. Since real dysfluent speech is unconstrained and word transcription is unknown (as shown in the "Human Transcription" in Figure \ref{UDM}, which is largely different from reference text), we develop hierarchical transcription methods. Firstly, we introduce an unconstrained forced aligner with a dynamic alignment search module to generate text-independent alignments. Secondly, we propose a Text Refresher that leverages the alignment input to refine the state-of-the-art Whisper~\cite{radford2022whisper} transcription. For the detection component, we employ \textit{2D-alignment} to automatically detect various phonetic and word-level dysfluency patterns, including repetition, missing, replacement, insertion, deletion, and irregular pauses. To further enhance performance, we curate a dysfluent dataset called VCTK$^{++}$ to boost the capacity of our unconstrained forced aligner (UFA). Experimental results demonstrate the effectiveness of our proposed framework in both dysfluent speech transcription and dysfluency pattern detection.

\section{Unconstrained Dysfluency Transcription}

\footnotetext{Audio sample for Fig.\ref{UDM}:\url{https://shorturl.at/rTW26}}
\subsection{Unconstrained Forced Aligner (UFA)}
\begin{figure}[h]
    \centering
    \includegraphics[height=6.4cm]{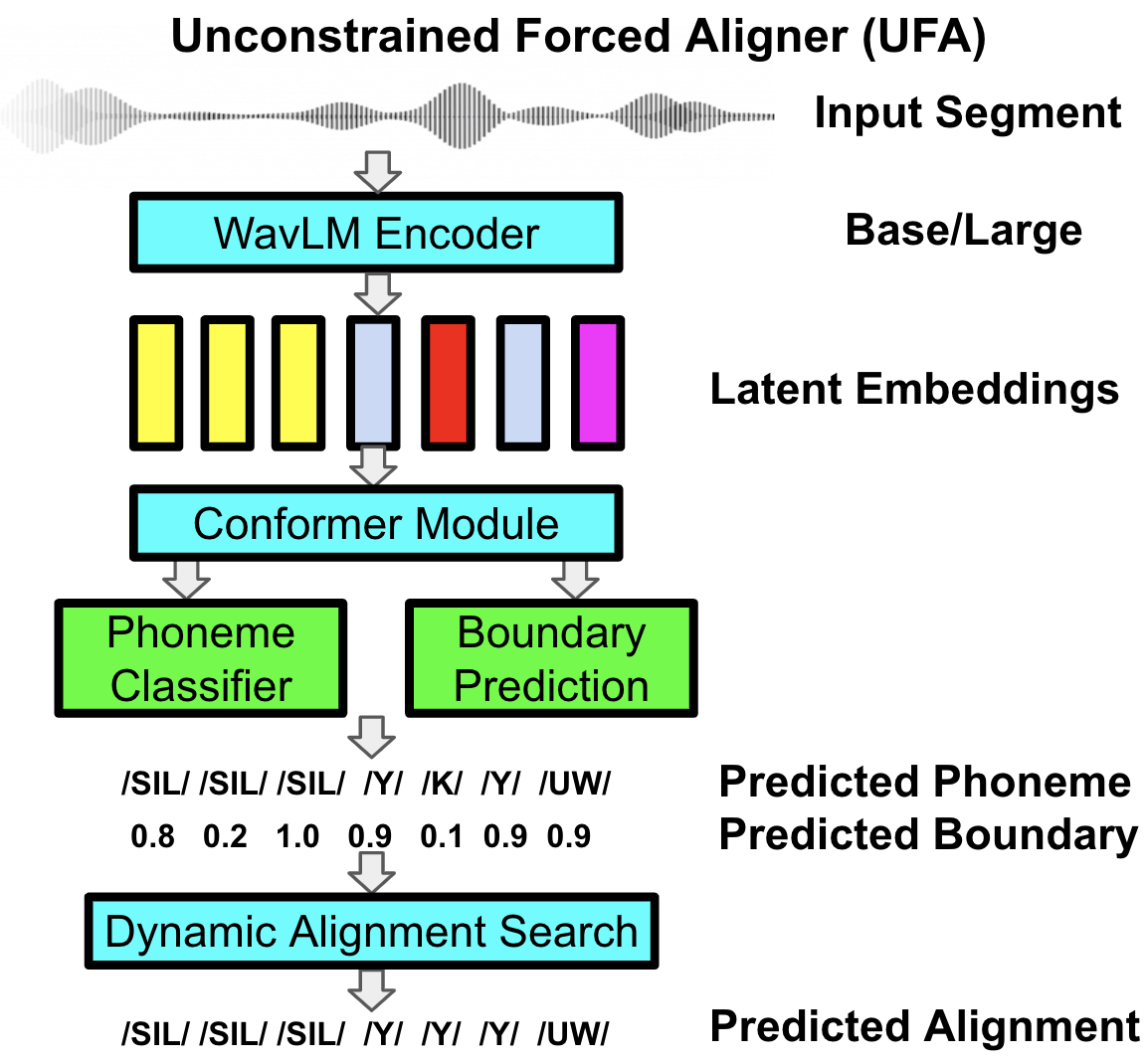}
    \caption{UFA Module}
    \label{ufa-fig}
\end{figure}
Unconstrained forced aligner predicts alignment with weak text supervision. As shown in Fig. \ref{ufa-fig}, a speech segment is passed into WavLM \cite{chen2022wavlm} encoder which generates latent representations. A conformer module \cite{gulati2020conformer} is followed to predict both alignment and boundary information. The alignment and boundary targets used in UFA are derived from the Montreal Forced Aligner (MFA) \cite{mcauliffe2017montrealmfa}. This integration with MFA allows UFA to operate with weak text supervision. During the inference stage, there is no need for text input, making the alignment process unconstrained. The conformer module comprises of four conformer~\cite{gulati2020conformer} encoder layers. The hidden size, number of attention heads, filter size, and dropout for each conformer layer are [1024, 4, 5, 0.1], [1024, 8, 3, 0.1], [1024, 8, 3, 0.1], [1024, 4, 3, 0.1] respectively. Two linear layers are simply applied as phoneme classifier and boundary predictor. For the phoneme classifier, UFA optimizes the softmax cross-entropy objective, while logistic regression is utilized for boundary prediction. Specifically, it predicts floating numbers between 0 (non-boundary) and 1 (boundary). 

\subsubsection{Dynamic Alignment Search} \label{DAS}
For the inference of dysfluent speech, such as aphasia speech, obtaining an accurate text transcription is often not achievable (as shown in the "Human Transcription" in Figure \ref{UDM}). Consequently, alignment should be decoded without text supervision. We propose a boundary-aware dynamic alignment search algorithm, which is the extension of Viterbi algorithm. Let us denote the phoneme logits as \textit{logits} $\in \mathbb{R}^{B,T,D}$, the boundary predictions as \textit{boundaries} $\in \mathbb{R}^{B,T}$, and the bi-gram phoneme language model as \textit{transition\_probs} $\in \mathbb{R}^{D,D}$, where $(B,T,D)$ represents the batch size, time steps, and phoneme dictionary size, respectively. The search algorithm is outlined in Algorithm 1.

It is worth noting that the transitions between consecutive phonemes near the boundary should be assigned lower importance to mitigate the risk of phoneme omissions. For instance, consider the correct alignment as SIL SIL SIL Y Y Y. In some cases, when the predicted probability for "Y" is low, there is a possibility that the prediction of "Y" might be overlooked due to the higher self-transition probability of SIL. Consequently, the final prediction could erroneously become SIL SIL SIL SIL SIL SIL. The bi-gram phoneme language model is derived by applying maximum likelihood estimation to the VCTK \cite{yamagishi2019cstr-vctk} forced alignment obtained from MFA \cite{mcauliffe2017montrealmfa}.
\begin{algorithm}
  \caption{Boundary-Aware Dynamic Alignment Search}\label{viterbi}
  \small
  \begin{algorithmic}[1]
    \Procedure{Decode}{\textit{logits}, \textit{boundaries}, \textit{transitional\_probs}}
      \State $B, T, D \gets$ shape of \textit{logits}
      \State Initialize \textit{trellis} and \textit{backpointers} 
      \For{$t$ in range($1$, $T$)}
        \For{$d$ in range($D$)}
          \State \textit{trellis}[:, $t$, $d$], \textit{backpointers}[:, $t$, $d$] $\gets$ \Call{\textbf{max\_argmax}}{\textit{trellis}[:, $t-1$, :] $+$ $(1-\textit{boundaries})$[:, $t$] $\times$ \textit{transition\_probs}[$d$, :]}
        \EndFor
      \EndFor
        \State Derive \textit{best\_path} from \textit{trellis} and \textit{backpointers}
      \State \textbf{return} \textit{best\_path}
    \EndProcedure
  \end{algorithmic}
\end{algorithm}
% \vspace{-5mm}
\subsection{ASR Transcription and Text-Refresher}
State-of-the-art ASR models \cite{radford2022whisper, zhang2023google-usm, pratap2023scaling} are commonly trained using a robust language model constraint, ensuring a high level of accuracy in transcribing dysfluent or disordered speech, thereby generating nearly perfect transcriptions. However, to perform word-level dysfluency analysis, it is necessary to introduce imperfections. In this study, we propose \textit{Text Refresher} to achieve this objective.

First, we obtain a perfect transcription using Whisper-large \cite{radford2022whisper}. We then obtain its corresponding phoneme transcription using CMU dictionary \cite{cmu-phoneme-dict}. Subsequently, in \textit{Text Refresher}, we perform Dynamic Time Warping (DTW) between the phoneme transcription of the Whisper output and the output of the Unsupervised Forced Aligner (UFA). Our primary focus is on identifying \textit{insertions} and \textit{deletions}. If a word (represented as a phoneme sequence) in the Whisper output does not align with the correct word (phoneme sequence) in the UFA output, we remove that word. For example, in the case illustrated in Figure \ref{UDM}, the word "to" is deleted. On the other hand, if a word (phoneme sequence) in the UFA output does not align with any word (phoneme sequence) in the Whisper output, we insert that word. Our observations indicate that in Aphasia speech, most word-level imperfections that Whisper cannot transcribe are primarily from deletions or insertions.

\subsection{Unconstrained Transcription Evaluation}

\subsubsection{Phonetic Transcription(Alignment) Evaluation}
 In order to evaluate how accurately the speech is transcribed at the frame level, we report \textbf{Micro F1 Score} and \textbf{Macro F1 Score} of phoneme transcription. Note that our F1 scores evaluate how many phonemes are correctly predicted. This is different from~\cite{strgar2022phoneme22SLT} which evaluates how many time steps are correctly predicted as phonetic boundaries. In order to evaluate the phoneme segmentation performance within our methods, we propose the duration-aware phoneme error rate (\textbf{dPER}). dPER extends Phoneme Error Rate (PER) by weighing each operation (substitution, insertion, deletion) with its duration. Denote $\hat{S}, \hat{I}, \hat{D}, \hat{C}$ as the weighted value of substitutions, insertions, deletions, and correct samples. Denote $p_i$ and $p_j$ as the current two phonemes we are comparing in the reference sequence and prediction sequence respectively. Denote $d(p_i)$ and $d(p_j)$ as their durations (number of repetitions). Whatever the error type is detected, we propose the following updating rule: $\hat{S} \rightarrow \hat{S}+ d(p_i)+d(p_j)$, $\hat{I}\rightarrow  \hat{I}+d(p_j)$, $\hat{D}\rightarrow \hat{D}+d(p_i)$, $\hat{C}\rightarrow \hat{C}+|d(p_i)-d(p_j)|$. The ultimate formula is: 
 \begin{equation}
 \resizebox{0.35\hsize}{!}{
     $\text{dPER}=\frac{\hat{S}+\hat{D}+\hat{I}}{\hat{S}+\hat{D}+\hat{C}}$
     }
 \end{equation} 
\subsubsection{Imperfect Word Transcription Evaluation}
In contrast to conventional ASR tasks, evaluating the performance in word-level dysfluency analysis requires the utilization of imperfect word targets. In this study, we employ manual word annotation of Aphasia speech as the target reference and report the \textit{imperfect Word Error Rate} (\textbf{iWER}). 
\section{Unconstrained Dysfluency Detection}
We develop rule-based methods for detecting time-accurate phonetic-level dysfluencies, including \textit{Phonetic Errors (Missing, Deletion, Replacement), Repetition, and Irregular Pause}. Additionally, our methods also cover word-level dysfluencies, including \textit{Missing, Insertion, Replacement, and Repetition}.
% \vspace{-4pt}
\subsection{Phonetic-Level Dysfluency Detection}
Phonetic-level dysfluency detection is empirically performed by comparing the predicted forced alignment with the reference ground truth text. The initial and crucial step in this process involves accurately \textbf{aligning the forced alignment with the reference text}. This alignment necessitates non-monotonicity whenever a phonetic dysfluency, such as repetition or insertion, occurs. Traditional aligners, such as Dynamic Time Warping (DTW) and MFA \cite{mcauliffe2017montrealmfa}, cannot handle non-monotonic cases and are inadequate for this purpose. 

\begin{figure}[h]
    \centering
    \includegraphics[height=7.2cm]{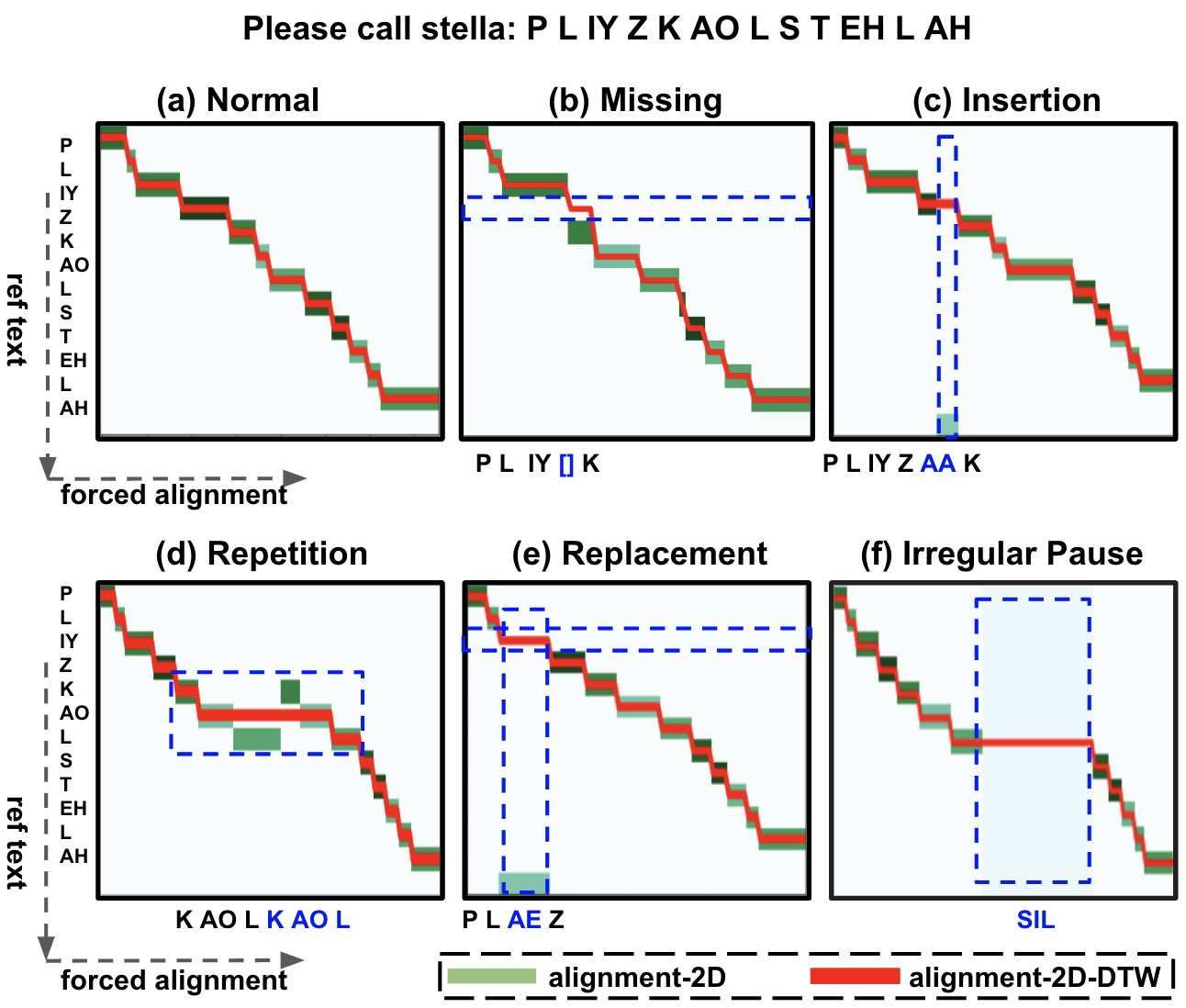}
    \caption{Phonetic-Level Dysfluency Detection. Audio samples can be found here. \protect\footnotemark}. 
    \label{detection-fig}
\end{figure}

To address this limitation, we extract the phoneme center embeddings from the phoneme classifier in the Unconstrained Forced Aligner (UFA) (Fig. \ref{ufa-fig}). By obtaining the phoneme embedding sequences for both the reference text and the forced alignment, we compute the dot product between these sequences. As a result, we generate a 2D similarity matrix that serves as the alignment representation. In the forced alignment, each phoneme may align with multiple occurrences of the same phoneme in the reference text, particularly when the reference text contains repeated phonemes. For instance, in the phrase "Please call Stella" represented as "P L IY Z K AO L S T EH L AH," each occurrence of "L" in the forced alignment aligns with all three "L" phonemes in the reference text. To ensure that only one phoneme in the reference aligns with the current phoneme in the forced alignment, we employ Viterbi search on the 2D similarity matrix. This process yields the final 2D alignment, denoted as \textbf{alignment-2d}, which is primarily monotonic. As illustrated in Figure \ref{UDM} and Figure \ref{detection-fig}, the alignment-2d is visualized through green plots, highlighting the relationship between the forced alignment and the reference text.

In addition to the alignment-2d, we also require a ground truth alignment-2d, which represents the expected alignment between the forced alignment of nearly perfect speech and the reference text. This ground truth alignment is strictly monotonic. To obtain it, we apply Dynamic Time Warping (DTW) between the forced alignment and the reference text, resulting in the alignment represented by the red plots in Figure \ref{detection-fig}. We denote this as \textbf{alignment-2D-DTW}. 

Finally, the detection of phonetic dysfluency becomes straightforward with the availability of the alignment-2D and alignment-2D-DTW. As illustrated in Figure \ref{detection-fig}, in the case of normal speech, these two alignments align perfectly with each other. However, if there is a significant decrease in the alignment-2D-DTW while lacking any intersection in the corresponding row, it indicates a \textbf{missing} phoneme, as depicted in Fig \ref{detection-fig} (b). If a row in alignment-2D-DTW encounters multiple columns in alignment-2D, and there are repeated phonemes present, it indicates a \textbf{repetition}. This is depicted in Figure \ref{detection-fig} (d). Conversely, if a row in alignment-2D-DTW already aligns with alignment-2D and simultaneously aligns with the surrounding column in alignment-2D, it signifies an \textbf{insertion}. This is illustrated in Figure \ref{detection-fig} (c). If a row in alignment-2D-DTW does not overlap with any horizontal regions in alignment-2D, but only overlaps with a single vertical block in alignment-2D, it is recognized as a \textbf{replacement}. This is depicted in Figure \ref{detection-fig} (e). Lastly, any pauses occurring within a complete sentence are identified as \textbf{irregular pauses}, as shown in Figure \ref{detection-fig} (f). It should be noted that within this rule-based detection framework, the precise timing of all five types of dysfluencies can be accurately identified with a resolution of 20ms.
% \vspace{-7pt}
\footnotetext{Audio sample for Fig.\ref{detection-fig}:\url{https://rb.gy/fmiv7}}
\vspace{-20pt}
\subsection{Word-level Dysfluency Detection}
We address \textit{missing}, \textit{insertion}, \textit{replacement}, and \textit{repetition} as part of our word-level dysfluency detection. To detect word-level dysfluency, we follow a similar methodology as phonetic-level dysfluency detection, which involves obtaining \textbf{alignment-2D} and \textbf{alignment-2D-DTW}.  However, in the case of word-level dysfluency detection, we do not utilize word embeddings. Instead, we employ perfect matching between the words in the reference and predicted texts, without the need for embedding dot product calculations. 
Duration, including silence, is not taken into account in this particular analysis since it is already incorporated in the phonetic component.
% \vspace{-5pt}
\subsection{Dysfluency Evaluation} \label{detect-eval-metric}
We conduct dysfluency evaluation on segments of Aphasia speech. In each Aphasia speech segment, manual annotations are made for all types of dysfluencies, including their accurate timing. For the evaluation of phonetic-level dysfluency, we report the \textbf{F1 score (Micro and Macro)} for dysfluency type identification. Additionally, we measure the accuracy of dysfluency detection in terms of time alignment. To achieve this, we calculate the Intersection over Union (IoU) between our predicted time boundaries and the ground truth boundaries. If the IoU is greater than 0.5, the dysfluency is identified as detected. We also report the F1 score for this matching evaluation, referred to as the \textbf{matching score}. For the evaluation of word-level dysfluency, we simply report the F1 score (Micro and Macro) without considering the timing aspects.

\section{Experiments}

\subsection{Datasets and Pre-processing} \label{datasets}
\paragraph*{VCTK~\cite{yamagishi2019cstr-vctk}} It is a multi-speaker accented corpus containing 44 hours of fluent speech. Following~\cite{lian2022robust-d-dsvae, lian2022towards-c-dsvae, lian2022utts}, we randomly select 90$\%$ of speakers as training set and the remaining as dev set. VCTK is used to train \textit{Unconstrained Forced Aligner}. 
% \vspace{-10pt}
\paragraph*{VCTK$^{++}$} For each waveform in VCTK and its forced alignment (from MFA~\cite{mcauliffe2017montrealmfa}), 
we applied simulations regarding the following stutter types. \textbf{(i) Repetitions}: Phonemes are randomly sampled within the waveform, appended by a variable-length sample of silence, and inserted into the original sound file. The silence sample is set to vary between 200ms and 500ms in multiples of 20 to match the framerate of the phoneme alignments. \textbf{(ii) Prolongations}: Phonemes are randomly selected, excluding phonemes that cannot be reasonably prolonged, such as hard consonants or silence tokens. The sound sample containing the phoneme is then stretched by a random factor anywhere from 5x to 10x using Waveform Similarity Overlap-Add (WSOLA)\cite{wsola}. The original phoneme is then replaced by the stretched variant in the waveform. \textbf{(iii) Blocks}: Phonemes are selected from a list of commonly blocked sounds, such as consonants or combinations of hard phonemes. With each simulation, we maintain the phoneme alignments such that the phoneme timestamps line up with the individual stutters, generating new alignments that act as ground truth for inference. See supplemental material for details. 
Here is an example of our augmented data.
\url{https://shorturl.at/aeGKR}
% \vspace{-3pt} 

\begin{table*}[h]
    \centering
    \setlength{\tabcolsep}{5pt}
    \renewcommand{\arraystretch}{1.2} 
     \resizebox{16cm}{!}{
    \begin{tabular}{l l c |c c c| c c c} 
    \toprule
    Method& WavLM Size&Training Data&Micro F1 ($\%$, $\uparrow$) & Macro F1 ($\%$, $\uparrow$) & dPER ($\%$, $\downarrow$) &Micro F1 ($\%$, $\uparrow$) & Macro F1 ($\%$, $\uparrow$) & dPER ($\%$, $\downarrow$)\\
     \hline
 \rowcolor{Gray} 
\multicolumn{3}{c}{}& \multicolumn{3}{c}{\textit{Buckeye Test Set}} & \multicolumn{3}{c}{\textit{VCTK++ Test Set}} \\
    WavLM-CTC-VAD&Large&None&50.1&47.3&86.9&48.8&45.7&88.0\\
    WavLM-CTC-MFA&Large&None&49.8&28.7&53.9&47.6&26.0&54.2\\
    UFA&Base&VCTK&68.9&55.6&53.3&78.8&59.5&53.4\\
    UFA&Base&VCTK+Buckeye&65.9&51.6&63.6&75.2&56.0&60.0\\
    UFA&Large&VCTK+Buckeye&70.3&55.0&46.2&80.7&66.4&45.8\\
    UFA&Large&VCTK&71.3&60.0&46.0&81.7&72.0&44.0\\
   \hspace{1em}-- Boundary-aware&Large&VCTK&68.9&52.0&49.9&78.4&62.9&47.8\\
UFA&Large&VCTK$^{++}$&\textbf{73.5}&\textbf{64.0}&\textbf{41.0}&\textbf{93.6}&\textbf{90.8}&\textbf{38.0}\\
\hspace{1em}-- Boundary-aware&Large&VCTK$^{++}$&71.0&63.7&44.3&91.1&90.0&42.1\\
    \bottomrule
    \end{tabular}}
    \caption{Phonetic transcription evaluation on two dysfluent corpora: Buckeye and VCTK++ test set.}
    \label{phn-transcription-eval}
\end{table*}

\begin{table}[h]
    \centering
    \setlength{\tabcolsep}{17pt}
    \renewcommand{\arraystretch}{1.1} % Adjust the row spacing
    \resizebox{7cm}{!}{
        \small
        \begin{tabular}{l c} 
            \toprule
            Method & iWER ($\%$, $\downarrow$)\\
            \hline
            Whisper-Large \cite{radford2022whisper} & 11.3 \\
            +Text Refresher & \textbf{9.7} \\
            \hspace{1em} +VCTK$^{++}$ & \textbf{9.2} \\
            \bottomrule
        \end{tabular}
    }
    \caption{Word Transcription Evaluation on Aphasia Speech}
    \label{word-trans-eval}
\end{table}
\paragraph*{Buckeye~\cite{pitt2005buckeye}} It contains over 40 hours of recordings from 40 native speakers of American English. The corpus contains quite a few portions of dysfluent speech with time-accurate annotation. We follow~\cite{strgar2022phoneme22SLT} to make the train/val/test split. The Buckeye corpus is utilized for training the \textit{Unconstrained Forced Aligner} and for \textit{Phonetic Transcription Evaluation}.
% \vspace{-3pt}
\paragraph*{Aphasia Speech} \label{Datasets}
From our clinical collaborators, our dysfluent data comprises ten participants diagnosed with Primary Progressive Aphasia (PPA), which exhibits various manifestations of dysfluencies. The dataset consists of audio recordings capturing interactions between patients and speech-language pathologists (SLPs). Our primary focus lies in the audio input of patients reading the Grandfather passage, resulting in approximately 20 minutes of speech data. The Aphasia speech dataset is employed for the evaluation of \textit{Imperfect Word Transcription} and \textit{Dysfluency Detection}.

\paragraph*{Phonetic Dictionary} \label{phndict}
We remove stress-aware phoneme labels (e.g. AE0, AE1$\rightarrow$AE). The phoneme dictionary adopted in this paper contains 39 monophones from CMU phoneme dictionary~\cite{cmu-phoneme-dict} along with one additional silence label. For Buckeye corpus, we manually translate the out-of-dictionary phonemes into CMU monophones. Here is the translation paradigm: AEN$\rightarrow$AE N, EYN$\rightarrow$EY N, IYN$\rightarrow$IY N, TQ$\rightarrow$T, IHN$\rightarrow$IH N, OWN$\rightarrow$OW N, NX$\rightarrow$N, EHN$\rightarrow$EH N, DX$\rightarrow$T, EN$\rightarrow$AH N, OYN$\rightarrow$OY N, EM$\rightarrow$EH M, ENG$\rightarrow$EH NG, EL$\rightarrow$EH L, AAN$\rightarrow$AA N, AHN$\rightarrow$AH N, AWN$\rightarrow$AW N. 
\vspace{-3pt}
\paragraph*{Audio Segmentation} \label{Pre-segmentation}
For VCTK, we train on the entire utterance without segmentation. For Buckeye data, we follow~\cite{strgar2022phoneme22SLT} to segment the long utterance by the ground truth transcription. We make sure that the beginning and ending silence length would be no longer than 3s, resulting in the length of all segments ranging from 2s to 17s. Different from~\cite{strgar2022phoneme22SLT}, we keep all silence labels but still remove the untranscriptable labels such as 'LAUGH', 'IVER', etc. For patient speech, we apply the online Silero VAD~\cite{Silero-VAD} with a default threshold of 0.5 to make the segments. We keep all of the silences and this results in the length of all segments ranging from 2s to 15s. All audio samples have a sampling rate of 16K Hz.

% \vspace{-5pt}
\subsection{Phonetic Transcription Experiments}
We train our unconstrained forced aligner (UFA) using three types of data: VCTK only, VCTK+Buckeye, and VCTK++. Additionally, we conduct an ablation study to examine the impact of the boundary-aware constraint in the dynamic search algorithm. This is achieved by removing the constraint from the search algorithm. Across all experiments, we utilize the same configuration settings, employing the Adam optimizer with an initial learning rate of 1e-3, which is decayed by 0.9 at each step. Each model converges after approximately 10 epochs, as determined by achieving a 90$\%$ phoneme classification accuracy on the development set.
\vspace{2pt}

Furthermore, we investigate two alternative forced aligners for comparison purposes: WavLM-CTC-VAD and WavLM-CTC-MFA. In WavLM-CTC-VAD, we combine the CTC phoneme alignment~\cite{kurzinger2020ctc-segmentation} obtained from WavLM-CTC~\cite{wavlm-ctc} with Voice Activity Detection (VAD) segmentation. By assigning blank tokens and incorporating silence segments identified using online Silero VAD~\cite{Silero-VAD}, we obtain a silence-aware transcription. The VAD threshold is set to the default value of 0.5, and the minimum and maximum speech durations are defined as 250ms and infinity, respectively. In WavLM-CTC-MFA, we employ the Montreal Forced Aligner (MFA)~\cite{mcauliffe2017montrealmfa} to derive silence-aware phoneme alignment. We utilize WavLM-CTC~\cite{wavlm-ctc} to generate the initial phoneme transcription, and we leverage a pre-trained English ARPA acoustic model. A pronunciation dictionary maps phonemes (as word-level items) to phonemes (as phonemic pronunciation breakdowns). The default beam size of 10 is applied for MFA. In the phoneme-to-phoneme dictionary, the parameters for each phoneme mapping include a pronunciation probability of 0.99, a silence probability of 0.05, and final silence and non-silence correction terms of 1.0. For both methods, no additional training data is needed.
Phonetic transcription results are shown in Table. \ref{phn-transcription-eval}. 
% \vspace{-20pt}
\subsection{Imperfect Word Transcription Experiments}
We utilize Whisper-large~\cite{radford2022whisper} as our ASR transcriber. We begin by presenting the results obtained directly from Whisper-large. Subsequently, we employ Text Refresher to refine the Whisper transcription and report the updated results. By default, Text Refresher incorporates the UFA-WavLM-Large-VCTK alignment. Additionally, for ablation purposes, we consider the UFA-WavLM-Large-VCTK$^{++}$ alignment as input, which demonstrated superior performance as indicated in Table \ref{phn-transcription-eval}. The comprehensive results are presented in Table \ref{word-trans-eval}.
% \vspace{-10pt}
\subsection{Dysfluency Detection}
\begin{table}[h]
    \centering
    \setlength{\tabcolsep}{3pt}
    \renewcommand{\arraystretch}{1.5} 
     \resizebox{8cm}{!}{
     \small
    \begin{tabular}{l c c c c} 
     \toprule
    Methods& F1 ($\%$, $\uparrow$) & MS ($\%$, $\uparrow$) & Human F1 ($\%$, $\uparrow$)& Human MS ($\%$, $\uparrow$)\\
     \hline
    UFA-VCTK&62.4&55.2&90.4&85.6\\
    UFA-VCTK$^{++}$&\textbf{64.5}&\textbf{60.2}&\textbf{90.6}&\textbf{86.0}\\
    \bottomrule
    \end{tabular}}
    \caption{Phonetic Dysfluency evaluation on Aphasia speech.}
    \label{phn-dysfluency-eval}
\end{table}
% \vspace{-1mm}
The preliminary experiments presented in Table \ref{phn-transcription-eval} indicate that both WavLM-CTC-VAD and WavLM-CTC-MFA do not exhibit significant improvements in phonetic transcription performance. Furthermore, the joint training of the VCTK and Buckeye corpora does not enhance the overall performance. Hence, we restrict our evaluation to two variants of the Unconstrained Forced Aligner (UFA): UFA-WavLM-Large-VCTK and UFA-WavLM-Large-VCTK$^{++}$. To assess the efficacy of our rule-based detection algorithm, we also perform manual detection using the predicted alignment from UFA and human-created targets. The results are presented in Table \ref{phn-dysfluency-eval} and Table \ref{word-dysfluency-eval}. MS is \textit{Matching Score}, as stated in Sec. \ref{detect-eval-metric}. 
% \vspace{-2mm}
\subsection{Results and Discussion}
\begin{table}[h]
    \centering
    \setlength{\tabcolsep}{10pt}
    \renewcommand{\arraystretch}{1.1} 
     \resizebox{8cm}{!}{
     \small
    \begin{tabular}{l c c} 
     \toprule
    Methods& F1 ($\%$, $\uparrow$) & Human F1 ($\%$, $\uparrow$)\\
     \hline
    Whisper-Large~\cite{radford2022whisper}&64.0&86.4\\
    +Text Refresher(VCTK)&66.8&88.0\\
    +Text Refresher(VCTK$^{++}$)&\textbf{68.4}&\textbf{89.1}\\
    \bottomrule
    \end{tabular}}
    \caption{Word Dysfluency evaluation on Aphasia speech.}
    \label{word-dysfluency-eval}
\end{table}
\subsubsection{Transcription Analysis}
We begin by examining the phonetic transcription results, as presented in Table \ref{phn-transcription-eval}. Both WavLM-CTC-VAD and WavLM-CTC-MFA demonstrate commendable zero-shot silence-aware phonetic transcription capabilities. However, their performance remains limited and is even inferior to the UFA trained with the WavLM base model. Interestingly, incorporating the Buckeye data during training does not yield any performance improvement. We hypothesize that the presence of noise in the Buckeye corpus, being a dysfluent dataset itself, hinders performance. Additionally, including the LibriSpeech dataset in VCTK training does not lead to performance enhancement. This suggests that UFA has already reached a certain limit in terms of data scalability. Consequently, the subsequent ablations and dysfluency detection experiments are conducted solely using UFA-WavLM-Large-VCTK.
During our ablation study, we consistently observed performance improvements by incorporating boundary prediction information in the dynamic alignment search, as described in Section \ref{DAS}. Moreover, our experiments on VCTK$^{++}$ consistently demonstrated enhanced performance compared to the original VCTK dataset, highlighting the robustness introduced by VCTK$^{++}$.
In terms of word transcription results, as shown in Table \ref{word-trans-eval}, we found that Whisper-Large exhibited the lowest performance due to its overpowering language modeling. However, with the introduction of Text Refresher and the incorporation of VCTK$^{++}$, we observed an improvement in the imperfect Word Error Rate (iWER), further boosting the overall performance.
% \vspace{-2mm}
\subsubsection{Dysfluency Analysis}
Since there is no previous work on hierarchical (word/phoneme) and fine-grained (time-accurate) dysfluency detection models like ours, we conducted ablation experiments to compare our proposed rule-based detection methods against ourselves. The results, as shown in Table \ref{phn-dysfluency-eval} and Table \ref{word-dysfluency-eval}, indicate impressive performance in terms of F1 scores and matching scores (MS), demonstrating the ability of our methods to accurately capture most dysfluencies. However, it is important to note that our methods still fall short of human detection performance, highlighting their inherent limitations.

\section{Conclusion and Limitations}
We propose an unconstrained dysfluency modeling approach that combines transcription and detection, which has been proven effective in both tasks. However, there are several limitations that should be addressed in future research. First, our detection experiments primarily focus on aphasia speech, which limits the generalizability. Future work should explore diverse and open-domain dysfluent datasets, which may lack manual annotations. Second, our approach relies on phoneme-level forced alignment as the key representation for detection. However, it is worth investigating alternative speech units, such as self-supervised discrete units~\cite{hsu2021hubert} and articulatory units~\cite{lian22bcsnmf, lian2023factor, inversion, wu23k_interspeech}, to improve alignment modeling. Third, while our detection method performs well on "easy" speech samples, it struggles with more challenging and abnormal speech patterns. Even in the current setting, the detection scores are still below human performance. Lastly, it is worth exploring the application of LLM-guided speech models~\cite{gong2023listen-llm-speech-1, rubenstein2023audiopalm-llm-speech-2} to advance dysfluency modeling in a prompt manner, which remains an open problem. Hopefully our proposed approach can serve as a promising paradigm for the field of dysfluency modeling. 

\section{Acknowledgement}
Thanks for support from UC Noyce Initiative, Society of Hellman Fellows, NIH/NIDCD and the Schwab Innovation fund.

%\vfill
%\pagebreak

% -------------------------------------------------------------------------
\bibliographystyle{IEEEbib}
\bibliography{main}

\end{document}